\documentclass[runningheads]{llncs}
\usepackage[T1]{fontenc}
\usepackage{amssymb}
\usepackage{amsmath}

\usepackage{graphicx}
\usepackage{multirow}
\usepackage{booktabs}
\usepackage{float}

\usepackage{hyperref}

\usepackage{bbding}

\usepackage{subfigure}

\begin{document}
\title{Knowledge-aware
Dual-side Attribute-enhanced Recommendation}
%

%
\author{Taotian Pang\inst{1} \and
Xingyu Lou\inst{2} \and
Fei Zhao\inst{1} \and
Zhen Wu\inst{1} \and 
Kuiyao Dong\inst{2} \and
Qiuying Peng\inst{2} \and
Yue Qi\inst{2} \and
Xinyu Dai\inst{1}}

\authorrunning{Pang et al.}
%

\institute{Nanjing University, Nanjing 210023, China \\
\email{\{pangtt,zhaof\}@smail.nju.edu.cn},\email{\{wuz,daixinyu\}@nju.edu.cn} \and
OPPO, Shenzhen, China\\
\email{\{louxingyu,dongkuiyao,pengqiuying,qiyue\}@oppo.com}
}

\maketitle              

\begin{abstract}
\textit{Knowledge-aware} recommendation methods (KGR) based on \textit{graph neural networks} (GNNs) and \textit{contrastive learning} (CL) have achieved promising performance. However, they fall short in modeling fine-grained user preferences and further fail to leverage the \textit{preference-attribute connection} to make predictions, leading to sub-optimal performance. To address the issue, we propose a method named \textit{\textbf{K}nowledge-aware \textbf{D}ual-side \textbf{A}ttribute-enhanced \textbf{R}ecommendation} (KDAR). Specifically, we build \textit{user preference representations} and \textit{attribute fusion representations} upon the attribute information in knowledge graphs, which are utilized to enhance \textit{collaborative filtering} (CF) based user and item representations, respectively. To discriminate the contribution of each attribute in these two types of attribute-based representations, a \textit{multi-level collaborative alignment contrasting} mechanism is proposed to align the importance of attributes with CF signals. Experimental results on four benchmark datasets demonstrate the superiority of KDAR over several state-of-the-art baselines. Further analyses verify the effectiveness of our method. The code of KDAR is released at: \href{https://github.com/TJTP/KDAR}{https://github.com/TJTP/KDAR}.

\end{abstract}
\section{Introduction}
\textit{Knowledge-aware} recommendation (KGR)~\cite{cke,kgat,kgin,mcrec} has gained growing attention in recent years. It employs a \textit{knowledge graph} (KG) as auxiliary information to address the data sparsity issue in traditional \textit{collaborative filtering} (CF) based recommendation \cite{cf,mf,ncf}. A KG is a heterogeneous graph consisting of different types of \textit{entities} and \textit{relations}, where nodes represent entities and edges denote relations. 

To utilize KGs effectively, a lot of efforts have been devoted, roughly divided into four categories: (1) \textbf{\textit{Knowledge graph embedding}} (KGE) based methods \cite{cke,ktup,kbe} learn embeddings of entities and relations based on transition constraints of graphs. However, the training pattern fails to model high-order connectivity explicitly. (2) \textbf{\textit{Path-based}} methods \cite{rkge,mcrec,hete-mf} select paths from user to item via entities and use them to train predictive models. However, the selection procedure is labor-intensive and requires domain expert knowledge. (3) \textbf{\textit{Graph neural network}} (GNN) based methods \cite{kgcn,kgat,ckan,kgin} are the mainstream nowadays. By stacking aggregation layers, GNNs can capture both first-order and high-order connectivity of graphs. Moreover, GNN-based methods are of great generality to different domains. However, the sparsity of supervision signals (\textit{i.e.}, CF signals) may impair the quality of representations. (4) \textbf{\textit{Contrastive learning}} (CL) is thus introduced to GNN-based methods \cite{kgcl,kgrec,kgic}. CL-based methods add self-supervised objects to provide auxiliary supervision, improving representation learning.

\begin{figure}[h]
    \centering
    \includegraphics[width=0.9\textwidth]{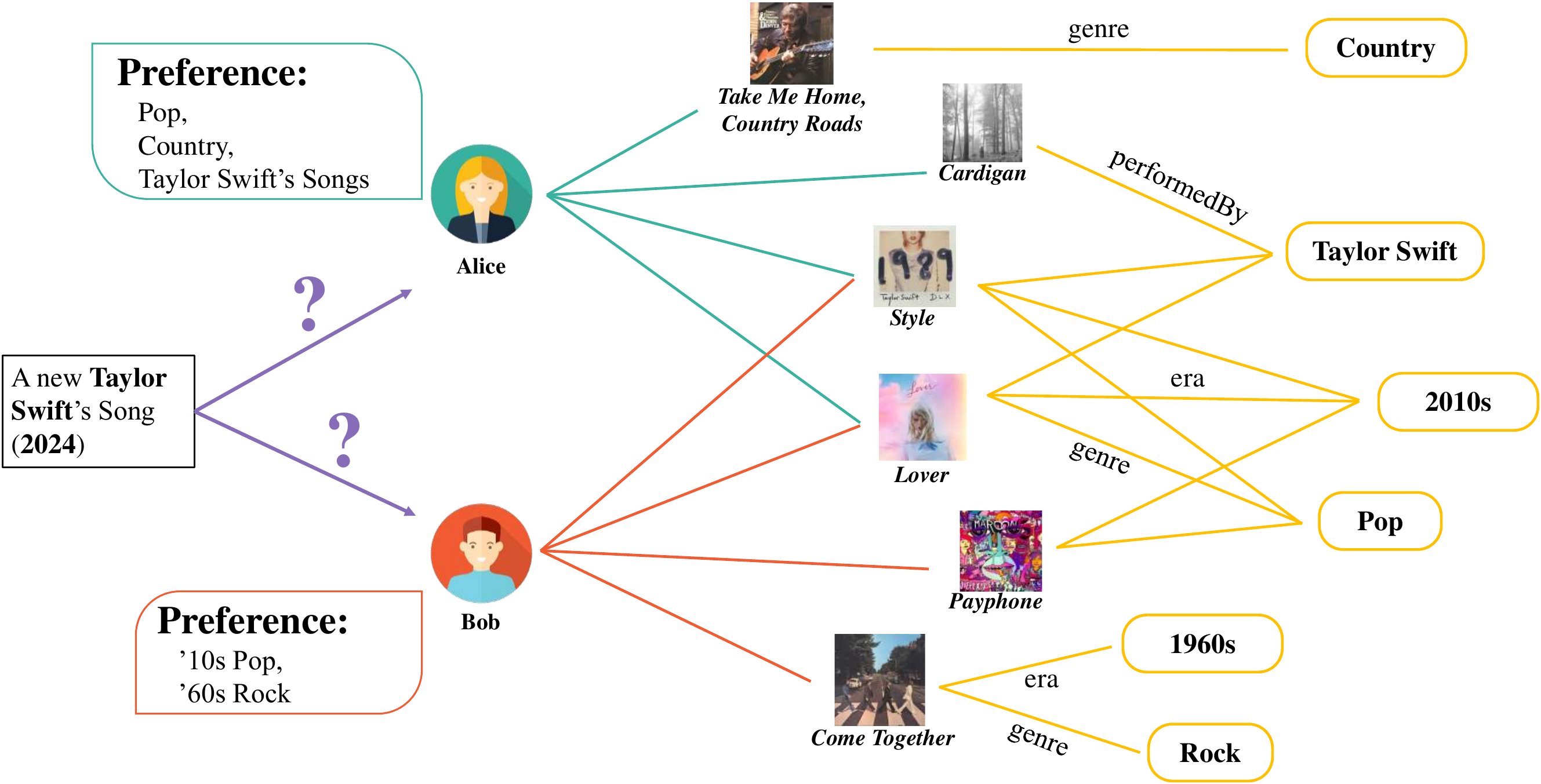}
    \caption{An example demonstrating the preference-attribute connection. In the figure, there are two users' historical interactions and preferences and a mini KG.}
    \label{fig:example}
\end{figure}
Despite their success, we argue that they fall short in modeling fine-grained user preferences from KGs and further fail to leverage the \textit{preference-attribute connection} to make recommendations, preventing them from achieving better performance. An observation of the real world is: User preferences are usually represented with item attributes. For example, in Fig. \ref{fig:example}, Alice's preference is ``Pop, Country music and Taylor Swift'', in which \textit{Pop}, \textit{Country} and \textit{Taylor Swift} are all terms of music attributes. The phenomenon indicates that: (1) User preferences can be modeled using attributes of his/her historical items. (2) The connection between user preferences and item attributes can be leveraged to predict users' interest in items more precisely. Taking the same example of Fig. \ref{fig:example}, from the perspective of CF, a Taylor Swift's song released in 2024 could be recommended to Alice and Bob because they both have listened some Taylor Swift's songs before. However, if considering their preferences and attributes of the song, it is not appropriate to push the song to Bob because he actually prefers \textit{'10s Pop} songs rather than \textit{Taylor Swift's songs}. This highlights the importance of modeling fine-grained user preferences and utilizing the preference-attribute connection to improve recommendation accuracy.

To tackle the above problems, we propose a method named \textit{\textbf{K}nowledge-aware \textbf{D}ual-side \textbf{A}ttribute-enhanced \textbf{R}ecommendation} (KDAR), which leverages item attribute information in KGs to enhance both user and item representations. Specifically, we build \textit{user preference representations} and \textit{attribute fusion representations} upon the attribute information in KGs, which are utilized to enhance CF-based user and item representations, respectively. To discriminate the contribution of each attribute in these two types of attribute-based representations, a \textit{multi-level collaborative alignment contrasting} mechanism is proposed to adjust the importance of attributes according to CF signals. Finally, the attribute-enhanced representations together with KG-based representations are concatenated to make predictions. The paradigm leverages the preference-attribute connection as well as collaborative and KG connectivity information for recommendation.

In summary, the contributions of this work are as follows:
\begin{itemize}
\item We discuss the connection between user preference and item attributes and how it can be leveraged to model fine-grained user preferences and improve recommendation performance.
\item We propose a new method named KDAR, which models user preferences from item attributes and leverages the preference-attribute connection to make recommendation. To distinguish the contribution of different attributes, we propose a multi-level collaborative alignment contrasting mechanism to adjust attribute weights.
\item We conduct empirical studies on four benchmark datasets to validate the effectiveness of the proposed method. The experimental results demonstrate the superiority of our method over the existed KGR methods.
\end{itemize}

\section{Preliminaries}
\textbf{Collaborative Data \& Graph}: Following formulation of mainstream works, we use $\mathcal{U}$ to denote the set of users and $\mathcal{I}$ to denote the set of items. The set $\mathcal{Y}^+=\{(u,i)|u\in \mathcal{U}, i\in \mathcal{I}\}$ represents observed interactions (click, rating, purchase, \textit{etc.}). Collaborative data can be represented as a bipartite graph, whic is named \textit{collaborative graph} (CG).

\noindent\textbf{Knowledge Graph}: A KG is defined as a set of triplets $\mathcal{G}=\{(h,r,t)| h, t\in\mathcal{V}, r\in \mathcal{R}\}$ , where the $\mathcal{V}$ means the set of entities and $\mathcal{R}$ means the set of relations. In each triplet, the relation $r$ describes the relation between head entity $h$ and tail entity $t$. For example, the triplet (\textit{Hey Jude}, performedBy, \textit{The Beatles}) shows the fact that the \textit{Hey Jude} is a song of  \textit{The Beatles}. The items are mapped to entities on KGs, \textit{i.e.}, $\mathcal{I}\subset \mathcal{V}$.

\noindent\textbf{Task description}: The task is using observed interactions $\mathcal{Y^+}$ and a knowledge graph $\mathcal{G}$ to predict the probability $\hat{y}_{ui}$ of a user $u$ to interact with an item $i$ that she/he has never interacted before.

\section{Methodology}
\begin{figure}[h]
    \centering
    \includegraphics[width=1.05\textwidth]{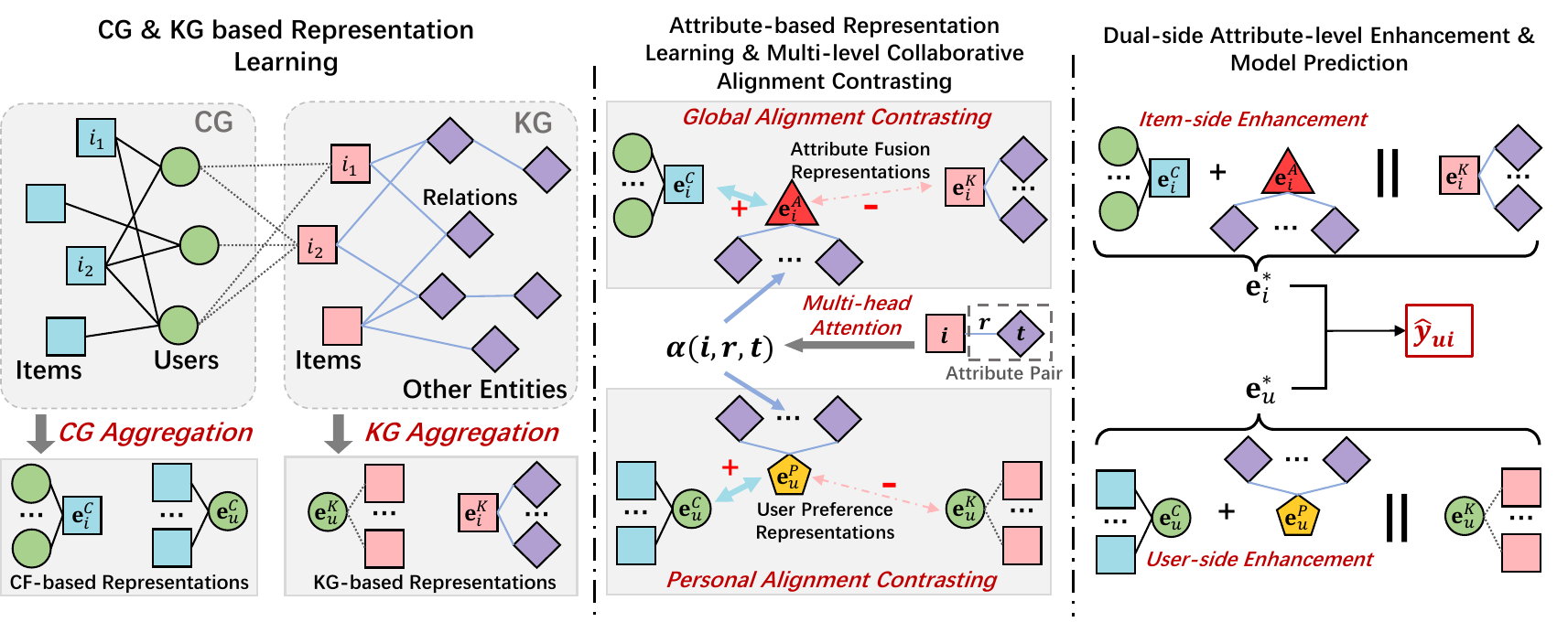}
    \caption{The overall framework of KDAR. The graph aggregation is formed on CG and KG respectively to obtain aggregated representations. The Attribute-based representations are constructed with multi-level collaborative alignment contrasting. Finally, dual-side attribute level enhancement is performed and the enhanced representations are concatenated with KG-based representations to make predictions.
    }
    \label{fig:model_arch}
\end{figure}

In this section, we present the details of the proposed KDAR. The overview of KDAR is shown in Fig. \ref{fig:model_arch}. KDAR consists of three main components: (1) CG and KG Representation Learning. (2) Attribute-based Representation Learning and Multi-level Collaborative Alignment Contrasting. (3) Dual-side Attribute-level 
Enhancement and Model Prediction.  

\subsection{CG \& KG Representation Learning}\label{sec:cg_kg_rep}
We first leverage GNNs to model the CF effect and the KG connectivity, which is achieved by performing the multi-hop aggregation on CGs and KGs respectively. CF signals serve as the golden standard to recommendation tasks and here we adopt the state-of-art GNN-based CF method LightGCN \cite{lightgcn} to model the CF effect:
\begin{equation}
    \mathbf{x}_i^{(l)}=\sum_{u\in\mathcal{N}_i^C}\frac{\mathbf{x}_u^{(l-1)}}{\sqrt{|\mathcal{N}_i^C||\mathcal{N}_u|}},\quad\mathbf{x}_u^{(l)}=\sum_{i\in\mathcal{N}_u}\frac{\mathbf{x}_i^{(l-1)}}{\sqrt{|\mathcal{N}_u||\mathcal{N}_i^C|}},\quad l=1,\cdots,L,
\end{equation} 
where $\mathbf{x}_i^{(l)},\mathbf{x}_u^{(l)}\in\mathbb{R}^d$ are representations of item $i$ and user $u$ on layer-$l$ respectively (Notice that $\mathbf{x}_\cdot^{(0)}$ means the ID embeddings of users or items), $L$ is the number of aggregation layers, $d$ is the dimensionality of vectors, $\mathcal{N}_u=\{i|(u,i)\in\mathcal{Y}^+\}$ represents the historical items of user $u$ and $\mathcal{N}_i^C=\{u|(u,i)\in\mathcal{Y}^+\}$ is the set of users who have interacted with item $i$. The representations of each layer are summed up to form the CF-based user representations $\mathbf{e}_u^{C}$ and item representations $\mathbf{e}_i^{C}$. 

KGs not only contain rich knowledge facts but also reveal heterogeneous connectivity among items and attributes, which can be leveraged to discover item correlations that are not expressed in CF signals, helping to improve recommendation. Following previous work \cite{kgin,kgrec}, we adopt an average relation-aware aggregation manner to model the KG connectivity. The aggregation layer is implemented as:
\begin{equation}
    \mathbf{e}_h^{(l)}=\frac{1}{|\mathcal{N}_h|}\sum_{(r,t)\in\mathcal{N}_h}\mathbf{e}_r\odot\mathbf{e}_t^{(l-1)},\quad l=1,\cdots,L,
\end{equation}
where $\mathbf{e}_\cdot^{(l)}\in\mathbb{R}^d$ denotes the representation of head or tail entities on layer-$l$ (Notice that $\mathbf{e}_\cdot^{(0)}$ means the ID embeddings of entities), $\mathcal{N}_h=\{(r,t)|(h,r,t)\in\mathcal{G}\}$ denotes the first-order connections of entity $h$, $\mathbf{e}_r$ is the embedding of relation $r$ and $\odot$ represents element-wise product operation. For each entity, the representation of each layer is summed up to form its KG-based representation $\mathbf{e}_h^{K}$.
For users, by iteratively aggregating representations of their historical items, we can implicitly inject KG facts and connectivity into user representations: 
\begin{equation}
    \mathbf{e}_u^{(l)}=\frac{1}{|\mathcal{N}_u|}\sum_{i\in\mathcal{N}_u}\mathbf{e}_i^{(l-1)},\quad l=1,\cdots,L.
\end{equation}
Similarly, we sum up user representations of each layer to form the KG-based user representations $\mathbf{e}_u^{K}$. In both CF-based and KG-based user representations, user preferences are modeled. However, they are implicitly modeled at a coarse-grained level, which are lack of precision. In next sections, we demonstrate how to model fine-grained user preferences and leverage them to make recommendations.

\subsection{Attribute-based Representation
Learning \& Multi-level Collaborative
Alignment Contrasting}\label{sec:align_contr}
In the real world, user preferences can be represented as combinations of item attributes and users choose items of which the attributes match with their preferences.  Based on these facts, we propose to leverage item attribute information
in KGs to model user preferences as well as enhance CF-based item representations. We also propose a special contrastive object to assign a proper weight to each attribute during representation learning.
\subsubsection{Attribute-based Representation Learning} Since an attribute describes \textit{the value} of \textit{a certain aspect} of items, it can be represented as a pair of a relation (corresponding to the \textit{aspect}) and an attribute entity (corresponding to the \textit{value}). Thus, the first-order connections of an item entity are the attributes of the item. To model each user's attribute-level preference, we collect attributes of each user's historical items and aggregate them. Moreover, since attributes are not of equal contributions to user preferences, an attention mechanism is incorporated to discriminate the importance of different attributes in representation learning:
\begin{equation}
    \mathbf{e}_{u}^P = \sum_{i\in\mathcal{N}_u}\sum_{(r,t)\in\mathcal{N}_i^K}\alpha(i,r,t)\mathbf{e}_r\odot\mathbf{e}_t^{(0)},
\end{equation}
where $\mathcal{N}_i^K=\{(r,t)|(i,r,t)\subset\mathcal{G}\}$ denotes the first-order connections of item $i$ in KGs and $\alpha(i,r,t)$ is the attentive weight for triplet $(i,r,t)$. Inspired by \cite{attneed}, the attention score $\alpha(i,r,t)$ is computed as:
\begin{equation}\label{eq:att_alpha}
\alpha(i,r,t)=\frac{\exp(\omega(i,r,t))}{\sum_{(i,r',t')\in\mathcal{N}_i}\exp(\omega(i,r',t'))},\quad
\omega(i,r,t)=\frac{\mathbf{e}_i\mathbf{W}_K\cdot(\mathbf{e}_r\odot\mathbf{e}_t\mathbf{W}_Q)^\top}{\sqrt{d}},
\end{equation}

where $\mathbf{W}_K,\mathbf{W}_Q\in\mathbb{R}^{d\times d}$ are trainable weights. 

To fully leverage the preference-attribute connection in recommendation, not only the user preferences should be modeled but also the attribute information should be emphasized at the item side. Thus we create an \textit{attribute fusion representation} for each item. Specifically, it is constructed by integrating attributes of each item:
\begin{equation}
    \mathbf{e}_{i}^{A} = \mathbf{e}_{i}^{(0)} + \sum_{(r,t)\in\mathcal{N}_i^K}\alpha(i,r,t)\mathbf{e}_r\odot\mathbf{e}_t^{(0)}.
\end{equation}
Here we use the same attention weights computed in Eq. (\ref{eq:att_alpha}) because the representative attributes in user preferences are likely to be important to items. 

\subsubsection{Multi-level
Collaborative Alignment Contrasting} To learn the attribute attention weights, apart from supervised signals, we further propose a \textit{multi-level collaborative alignment contrasting} mechanism, which aligns the importance of attributes according to the semantics of CF signals. Specifically, it consists of \textit{global alignment contrasting} (GAC) and \textit{personal alignment contrasting} (PAC), based on which it aligns the importance of attributes with both global and personal user behavior patterns contained in collaborative data, respectively.

From global views, for each item, the group of users who have interacted with it focus on some common attributes of it while other attributes of it have less influence on users' decision. Thus the GAC is designed to adjust the weights of attributes to reflect this global user behavior pattern. Specifically, we force the attribute fusion representations to align with CF-based item representations, which contain information of interacted users:
\begin{equation}
    \mathcal{L}_{GAC}=\sum_{i\in\mathcal{V}}-\log\frac{\exp(s(\mathbf{e}^A_i,\mathbf{e}_i^C)/\tau)}{\exp(s(\mathbf{e}^A_i,\mathbf{e}_i^C)/\tau)+\exp(s(\mathbf{e}^A_i,\mathbf{e}_i^K)/\tau)},
\end{equation}
where $\tau$ is the temperature parameter and $s(\cdot,\cdot)$ is the cosine similarity. In the contrastive object $\mathcal{L}_{GAC}$, we use KG-based item representations as negative samples. From personal views, for each user, his or her purchasing habit is dominated by a group of attributes. In PAC, we use semantics of this personal behavior pattern to adjust the weights by aligning user preference representations with CF-based user representations:
\begin{equation}
    \mathcal{L}_{PAC}=\sum_{u\in\mathcal{U}}-\log\frac{\exp(s(\mathbf{e}^P_u,\mathbf{e}_u^C)/\tau)}{\exp(s(\mathbf{e}^P_u,\mathbf{e}_u^C)/\tau)+\exp(s(\mathbf{e}^P_u,\mathbf{e}_u^K)/\tau)}.
\end{equation}

\subsection{Dual-side Attribute-level  Enhancement \& Model Prediction}
\label{sec:dual_enhan_mp}
After obtaining two types of attribute-based representations, we use them to enhance the CF-based representations and then the preference-attribute connection could be leveraged in making recommendations. The enhancement is implemented as follows:
\begin{equation}
    \mathbf{e}_u^E=(\mathbf{e}_u^C+\mathbf{e}_u^P) / 2,\quad
    \mathbf{e}_i^E=(\mathbf{e}_i^C+\mathbf{e}_i^A) / 2.
\end{equation}
We concatenate these attribute-enhanced representations and KG-based representations to construct final representations $\mathbf{e}^\ast_u$ and $\mathbf{e}^\ast_i$ respectively:
\begin{equation}
    \mathbf{e}^\ast_u=\mathbf{e}_u^E||\mathbf{e}_u^K,\qquad\mathbf{e}^\ast_i=\mathbf{e}_i^E||\mathbf{e}_i^K,
\end{equation}
where $||$ is the concatenation operator. The prediction of the probability that user $u$ would interact with item $i$ is computed by inner product:
\begin{equation}
    \hat{y}_{ui}={\mathbf{e}^\ast_u}^\top\mathbf{e}^\ast_i.
\end{equation}
\subsection{Model Optimization}
To train the model to make correct predictions, we adopt the pairwise BPR loss \cite{bpr} to optimize model parameters: 
\begin{equation}
    \mathcal{L}_{\text{BPR}}=\sum_{(u,i,j)\in\mathcal{Y}}-\ln\sigma(\hat{y}_{ui}-\hat{y}_{uj}),
\end{equation}
where $\mathcal{Y}=\{(u,i,j)|(u,i)\in\mathcal{Y}^+,(u,j)\in\mathcal{Y}^-\}$, and $\sigma(\cdot)$ denotes the sigmoid function. The $\mathcal{Y}^-$ is the set of unobserved interactions. In the two contrastive objects, CF-based representations are served as positive samples to provide semantics of CF signals to attribute weights. Thus, it is important to ensure the CF effect is properly modeled in the CF-based representations. To achieve the goal, we compute an extra BPR loss $\mathcal{L}_{\text{BPR}}^C$ based on them to optimize them especially. The overall object function is as follow:
\begin{equation}
    \mathcal{L}=\mathcal{L}_{\text{BPR}}+\lambda_1\mathcal{L}_{\text{BPR}}^C+\lambda_2(\mathcal{L}_{GAC}+\mathcal{L}_{PAC}) + \lambda_3\|\Theta\|^2_2,
\end{equation}
where $\|\Theta\|^2_2$ is the $L_2$ regularization term and $\lambda_\cdot$ are the hyperparameters controlling the weight of each object function. 

\section{Experiments}
In this section, we conduct empirical studies to demonstrate the effectiveness of the proposed KDAR. We aim to answer the following research questions:
\begin{itemize}
    \item \textbf{RQ1:} How does KDAR perform, comparing with the state-of-the-art knowledge-aware methods and other baselines?
    \item \textbf{RQ2:} What is the contribution of key components to the model performance?
    \item \textbf{RQ3:} How do different hyperparameters settings influence KDAR?
    \item \textbf{RQ4:} Can KDAR perform well on some specific situations such as cold-start and long-tail item recommendation?
\end{itemize}
\subsection{Experimental Settings}
\subsubsection{Dataset Description} We utilize four benchmark datasets in our experiments for music, book, movie and restaurant recommendation respectively:
(1) \textbf{Last.FM}\footnote{https://grouplens.org/datasets/hetrec-2011/} contains music listening records of a set of around 2 thousands users collecting from Last.FM online music system. (2) \textbf{Amazon Book}\footnote{https://jmcauley.ucsd.edu/data/amazon/} is collected from the review data of book category on Amazon website. (3) \textbf{MovieLens-20M}\footnote{https://grouplens.org/datasets/movielens/} consists of around 20 million ratings from the MovieLens website.(4) \textbf{Yelp2018}\footnote{https://www.yelp.com/dataset} is collected from 2018 edition of the Yelp challenge, in which the items are restaurants and bars. We use the original data released by KGCN \cite{kgcn} and KGAT \cite{kgat}. To ensure the quality of the data, we adopt the 5-core settings, \textit{i.e.}, retaining users with at least five interaction records. The statistics of datasets are summarized in Table \ref{tab:stat_data}.

\begin{table}[h]
\begin{center}
  \caption{The statistics of four datasets.}
  \label{tab:stat_data}
  \begin{tabular}{c|l|r|r|r|r}
    \toprule
    & & Last.FM & Amazon Book & MovieLens-20M & Yelp2018\\
    \hline
    \multirow{3}{*}{User-Item Interaction} & \#Users & 1,815 & 70,679 & 133,785 & 45,919\\
    & \#Items & 3,846 & 24,915 & 16,953 & 45,538\\
    & \#Interactions & 20,996 & 846,098 & 6,737,656 & 1,183,610\\
    \hline
    \multirow{3}{*}{Knowledge Graph} & \#Entities & 9,366 & 113,487 & 102,569 & 136,499\\
    & \#Relations & 60 & 39 & 32 & 42 \\
    & \#Triplets & 15,518 & 2,557,746 & 499,474 & 1,853,704\\
    \bottomrule
  \end{tabular}
\end{center}
\end{table}

\subsubsection{Evaluation Metrics} Following KGAT \cite{kgat} and KGIN \cite{kgin}, we adopt the all-ranking strategy in evaluation phase. We randomly split each dataset into training and test sets with the ratio of $4:1$. All items that a user has never interacted with before are considered as negative. We evaluate our model on three metrics: Recall@$K$, NDCG@$K$ 
and AUC . We vary the $K$ in range of $\{5, 10, 20, 50, 100\}$ and mainly focus on results of Recall@$20$ and NDCG@$20$. The reported results are the mean of all users in test set.

\subsubsection{Baselines} We compare our proposed KDAR with the following methods that cover CF-based (MF, LightGCN), KGE-based (CKE), GNN-based (KGCN, KGAT, CKAN, and KGIN) and CL-based (KGCL, KGRec) methods:
\begin{itemize}
    \item \textbf{MF} \cite{mf} factorizes user-item interaction matrices to make predictions.
    \item \textbf{LightGCN} \cite{lightgcn} is the state-of-art GNN-based CF methods.
    \item \textbf{CKE} \cite{cke} is a KGE-based method, which incorporates embeddings derived from TransR \cite{transr} to MF framework.
    \item \textbf{KGCN} \cite{kgcn} applies non-spectral GCN to KG to discover structure information and captures users’ potential long-distance interests.
    \item \textbf{KGAT} \cite{kgat} adopts attention mechanism into collaborative knowledge graph aggregation.
    \item \textbf{CKAN} \cite{ckan} proposes heterogeneous propagation to encode both collaborative information and KG information.
    \item \textbf{KGIN} \cite{kgin} is the state-of-the-art GNN-based method which models the user-item interaction relation at user intents level.
    \item \textbf{KGCL} \cite{kgcl} introduces graph contrastive learning to denoise KGs. The KG contrastive signals are further utilized to denoise the user-item graph.
    \item \textbf{KGRec} \cite{kgrec} is the latest state-of-the-art CL-based methods which proposes to learn the rationales in KG connections by contrasting the CF signals with KG representations.
\end{itemize}
\subsubsection{Parameter Settings} The size of ID embeddings $d$ is fixed as $64$, the batch size is set as $2048$. The number of GNN layers $L$ is set as $3$ and the temperature $\tau$ is set as $1$. The learning rate is set as $10^{-4}$. We compare the performance of KDAR with baselines using their official codes.

\subsection{Performance Comparison(RQ1)}

\begin{figure}
  \centering
  \subfigure[Last.FM]{\includegraphics[width=0.24\textwidth]{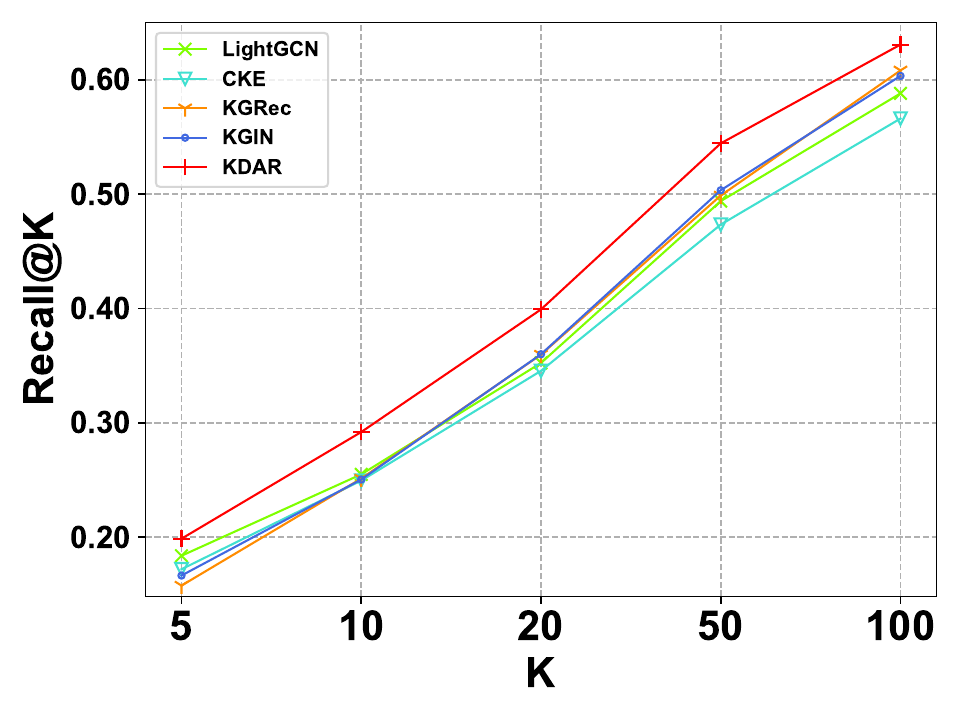}}
  \subfigure[Amazon Book]{\includegraphics[width=0.24\textwidth]{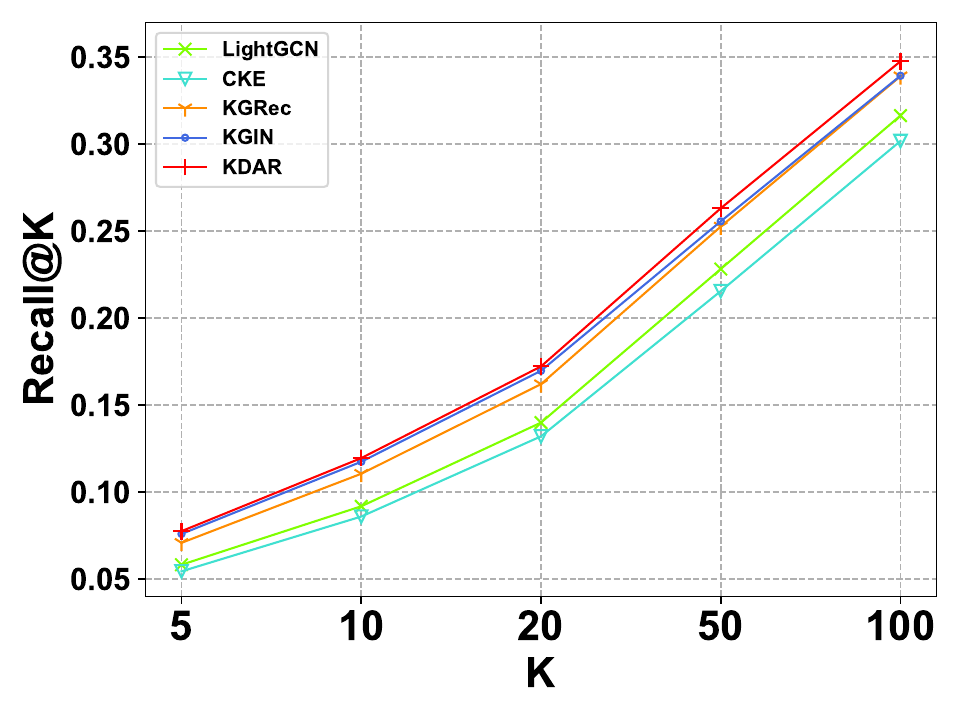}}
  \subfigure[MovieLens-20M]{\includegraphics[width=0.24\textwidth]{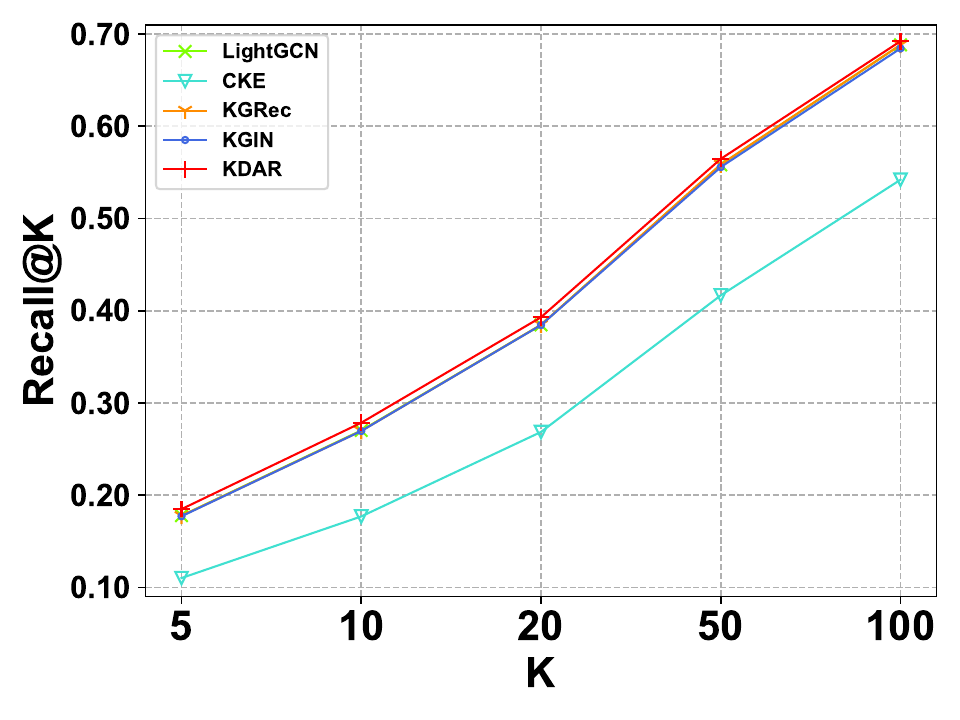}}
  \subfigure[Yelp2018]{\includegraphics[width=0.24\textwidth]{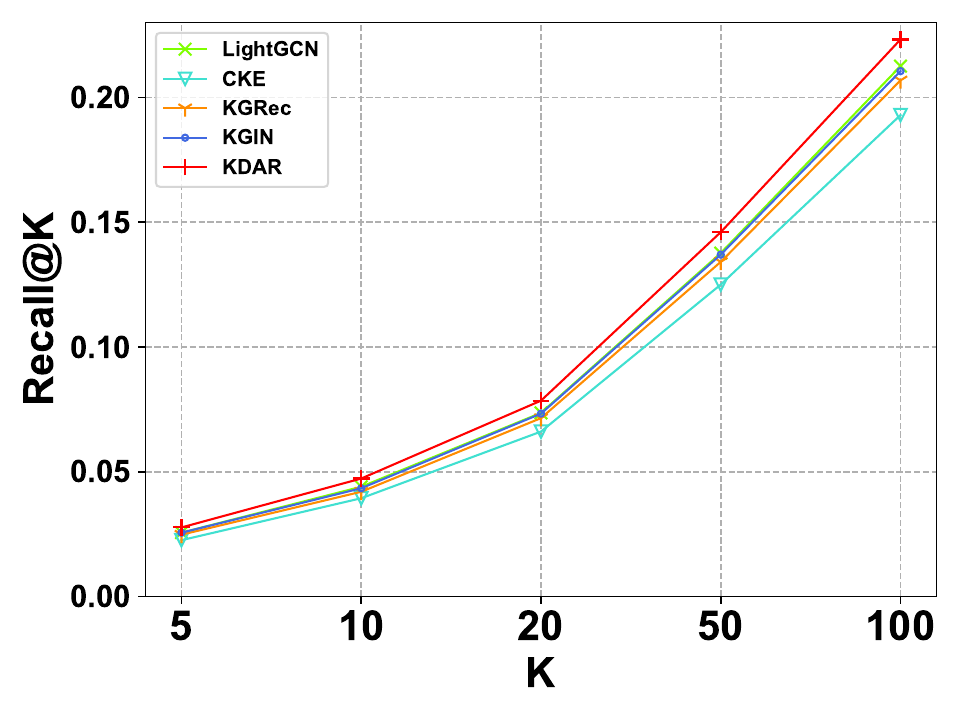}}
  
  \caption{The curves of Recall@$K$ on four datasets.}
  \label{fig:recallcurve}
\end{figure}

The results of our proposed KDAR and other baselines are presented in Table \ref{tab:perform_sota} (With AUC, Recall@$20$ and NDCG@$20$). The results of the best performing method are shown in bold while the results of the second strongest method are underlined. Fig. \ref{fig:recallcurve} shows curves of Recall@$K$ of our model and four baselines. We have following findings:

\begin{table}[h]
    \centering
  \caption{Overall performance comparison}
  \label{tab:perform_sota}
  \scalebox{0.8}{
  \begin{tabular}{c|ccc|ccc|ccc|ccc}
    \toprule[1pt]
    & \multicolumn{3}{c|}{Last.FM} & \multicolumn{3}{c|}{Amazon Book} & \multicolumn{3}{c|}{MovieLens-20M} & \multicolumn{3}{c}{Yelp2018}\\
    & AUC & Recall & NDCG & AUC & Recall & NDCG & AUC & Recall & NDCG & AUC & Recall & NDCG\\
    \hline
    MF & 0.8678 & 0.3470 & 0.2018 & 0.9090 & 0.1324 & 0.0687 & 0.9788 & 0.2696 & 0.2075 & 0.9535 & 0.0661 & 0.0420\\
    LightGCN & 0.8925 & 0.3523 & 0.2048 & 0.9121 & 0.1400 & 0.0728 & 0.9869 & \underline{0.3846} & 0.3102 & 0.9547 & \underline{0.0736} & \underline{0.0477}\\
    \hline
    CKE & 0.8672 & 0.3454 & 0.1996 & 0.9077 & 0.1321 & 0.0684 & 0.9788 & 0.2685 & 0.2066 & 0.9533 & 0.0661 & 0.0421\\
    \hline
    KGCN & 0.8352 & 0.2600 & 0.1200 & 0.7678 & 0.0389 & 0.0141 & 0.9744 & 0.2030 & 0.1489 & 0.7336 & 0.0238 & 0.0131\\
    KGAT & 0.8832 & 0.3363 & 0.1847 & 0.9041 & 0.1520 & 0.0808 & 0.9866 & 0.3813 & 0.3016 & 0.9482 & 0.0700 & 0.0449\\
    CKAN & 0.8646 & 0.2565 & 0.1350 & 0.8709 & 0.0764 & 0.0385 & 0.9776 & 0.1891 & 0.1517 & 0.8930 & 0.0296 & 0.0184\\
    KGIN & 0.8883 & 0.3415 & 0.1887 & 0.9142 & \underline{0.1698} & \underline{0.0921} & \underline{0.9871} & 0.3844 & 0.3092 & \underline{0.9573} & 0.0733 & 0.0476\\
    \hline
    KGCL & 0.8807 & \underline{0.3831} & \textbf{0.2311} & 0.8859 & 0.1596 & 0.0847 & 0.9662 & 0.3329 & 0.2636 & 0.9006 & 0.0701 & 0.0460\\
    KGRec & \underline{0.8989} & 0.3596 & 0.1906 & \underline{0.9172} & 0.1621 & 0.0864 & \textbf{0.9874} & \underline{0.3846} & 0.3084 & 0.9525 & 0.0715 & 0.0462\\
    \hline
    \textbf{KDAR} & \textbf{0.9084} & \textbf{0.3991} & \underline{0.2236} & \textbf{0.9162} & \textbf{0.1723} & \textbf{0.0929} & \underline{0.9871} & \textbf{0.3929} & \textbf{0.3190} & \textbf{0.9594} & \textbf{0.0785} & \textbf{0.0509}\\
    \bottomrule[1pt]
  \end{tabular}}
\end{table}

\begin{itemize}
    \item Our proposed KDAR outperforms all baselines on Recall@20 metric of four datasets. For other two metrics, KDAR outperforms all baselines in most cases except achieves the second best on the NDCG@20 of Last.FM dataset and the AUC of MovieLens-20M dataset. The results demonstrate the superiority and stability of our method over other baselines because the ranks of all baselines on three metrics are very unstable across different datasets, indicating they fail to model the general features of different domains. Even the latest state-of-the-art baseline KGRec is beaten by CF-based methods on some datasets. We attribute the superiority and stability of KDAR to leveraging the widely existed preference-attribute connection to make recommendations in different domains. 
    \item From the Recall@$K$ curves in Fig. \ref{fig:recallcurve}, we find our model consistently outperforms other baselines from $K=5$ to $K=100$ on four datasets. Especially, on Amazon Book and Yelp2018 dataset, our method extends the lead when $K\geq50$ while it uniformly takes the lead on Last.FM and MovieLens-20M dataset from $K=5$ to $K=100$.
    \item Some GNN-based methods are outperformed by KGE-based methods, which suggests that GNNs may introduce noises from their high-order aggregation.
    \item Comparing CL-based methods with GNN-based methods, we find that the former is not always better than the latter, indicating that the CL objects may conflict with the recommendation tasks, thus impairing the performance. 
    \item As the methods without KGs, MF and LightGCN still defeat some knowledge-aware methods on some metrics respectively, indicating that some KGR methods fail to incorporate and leverage KG information effectively. 
\end{itemize}

\subsection{Ablation Study(RQ2)}
In order to evaluate contributions of key components to the performance, we conduct ablation studies with four variants of KDAR: (1)\textbf{w/o Enhancement}, which removes attribute-level enhancement module (\textit{i.e}, no user preferences and attribute fusion representations). (2) \textbf{w/o ATTN}, which assigns equal importance to each attributes. (3) \textbf{w/o CL}, which abandons all CL objects. (4) \textbf{w/o CG}, which relies on only KG representations to recommend. 
\begin{table}[h]
\centering
  \caption{The ablation results of KDAR with different variants.}
  \label{tab:variants}
  \scalebox{0.75}{
  \begin{tabular}{l|ccc|ccc|ccc|ccc}
    \toprule[1pt]
    & \multicolumn{3}{c|}{Last.FM} & \multicolumn{3}{c|}{Amazon Book} & \multicolumn{3}{c|}{MovieLens-20M} & \multicolumn{3}{c}{Yelp2018}\\
    & AUC & Recall & NDCG & AUC & Recall & NDCG & AUC & Recall & NDCG & AUC & Recall & NDCG\\
    \hline
    \textbf{KDAR} & \textbf{0.9084} & \textbf{0.3991} & \textbf{0.2236} & \textbf{0.9162} & \textbf{0.1723} & \textbf{0.0929} & 0.9871 & \textbf{0.3929} & \textbf{0.3190} & 0.9594 & 0.0785 & 0.0509\\
    \hline
    w/o Enhancement & 0.9006 & 0.3589 & 0.2002 & 0.9144 & 0.1700 & 0.0919 & 0.9870 & 0.3898 & 0.3156 & 0.9585 & \textbf{0.0789} & \textbf{0.0511}\\
    w/o ATTN & 0.8968 & 0.3691 & 0.2078 & 0.9157 & 0.1719 & 0.0929 & 0.9870 & 0.3914 & 0.3183 & \textbf{0.9598} & 0.0783 & 0.0506\\
    w/o CL & 0.8973 & 0.3469 & 0.1942 & 0.9160 & 0.1723 & 0.0925 & 0.9871 & 0.3910 & 0.3188 & 0.9594 & 0.0785 & 0.0509\\
    w/o CG & 0.8911 & 0.3394 & 0.1879 & 0.9128 & 0.1656 & 0.0898 & \textbf{0.9872} & 0.3918 & 0.3152 & 0.9558 & 0.0733 & 0.0479\\
    \bottomrule[1pt]
  \end{tabular}}
\end{table}

The results of the four variants are reported in Table \ref{tab:variants} and we have following observations: (1) The performance decreases most when CGs are removed, indicating that the modeling CF effect is crucial to recommendation tasks. (2) The contributions of attention mechanism and contrastive objects to improve the performance vary across different datasets. (3) Removing the enhancement component impairs the performance on most datasets, suggesting that explicitly leveraging the preference-attribute connection is beneficial for recommendation.

\subsection{Hyperparameters Study(RQ3)}
\begin{figure}[h]
  \centering
  \subfigure[Last.FM]{\includegraphics[width=0.24\textwidth]{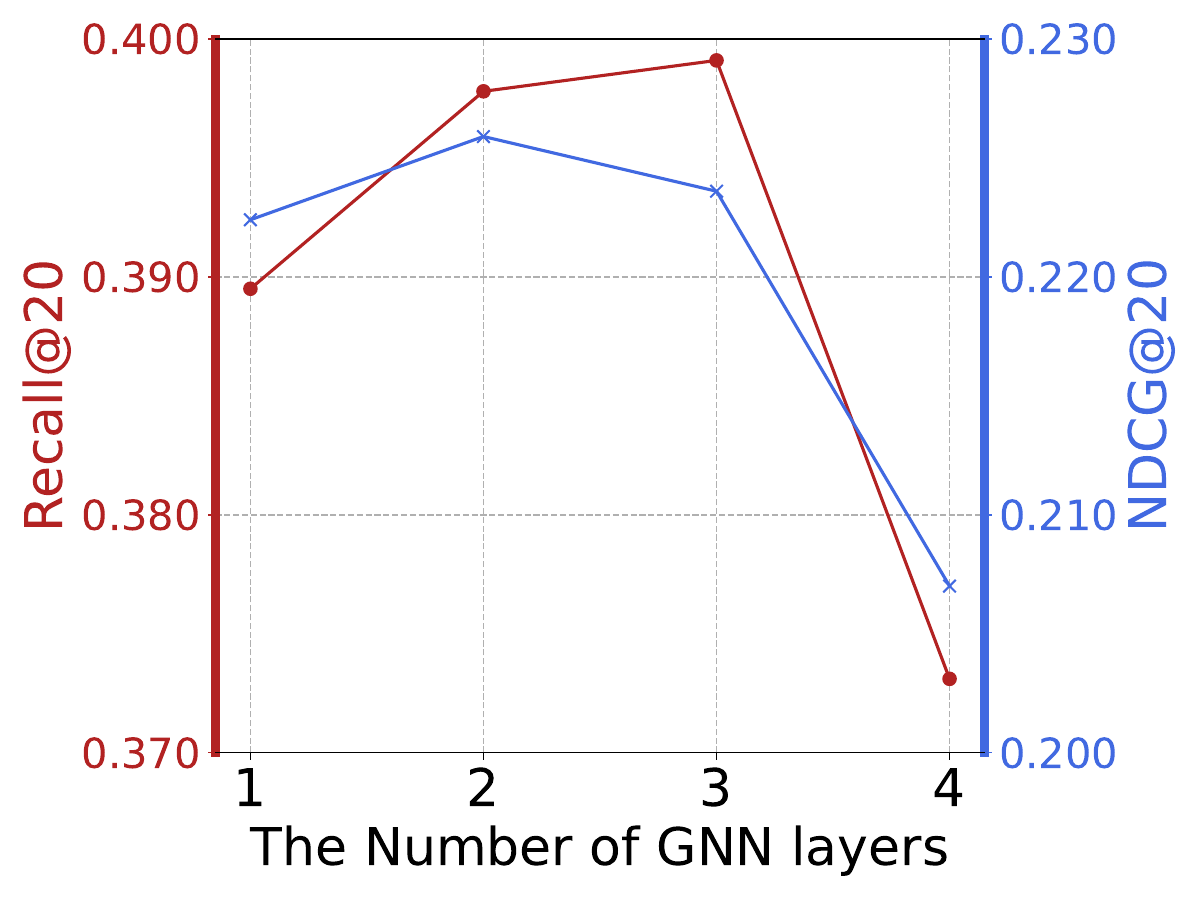}}
  \subfigure[Amazon Book]{\includegraphics[width=0.24\textwidth]{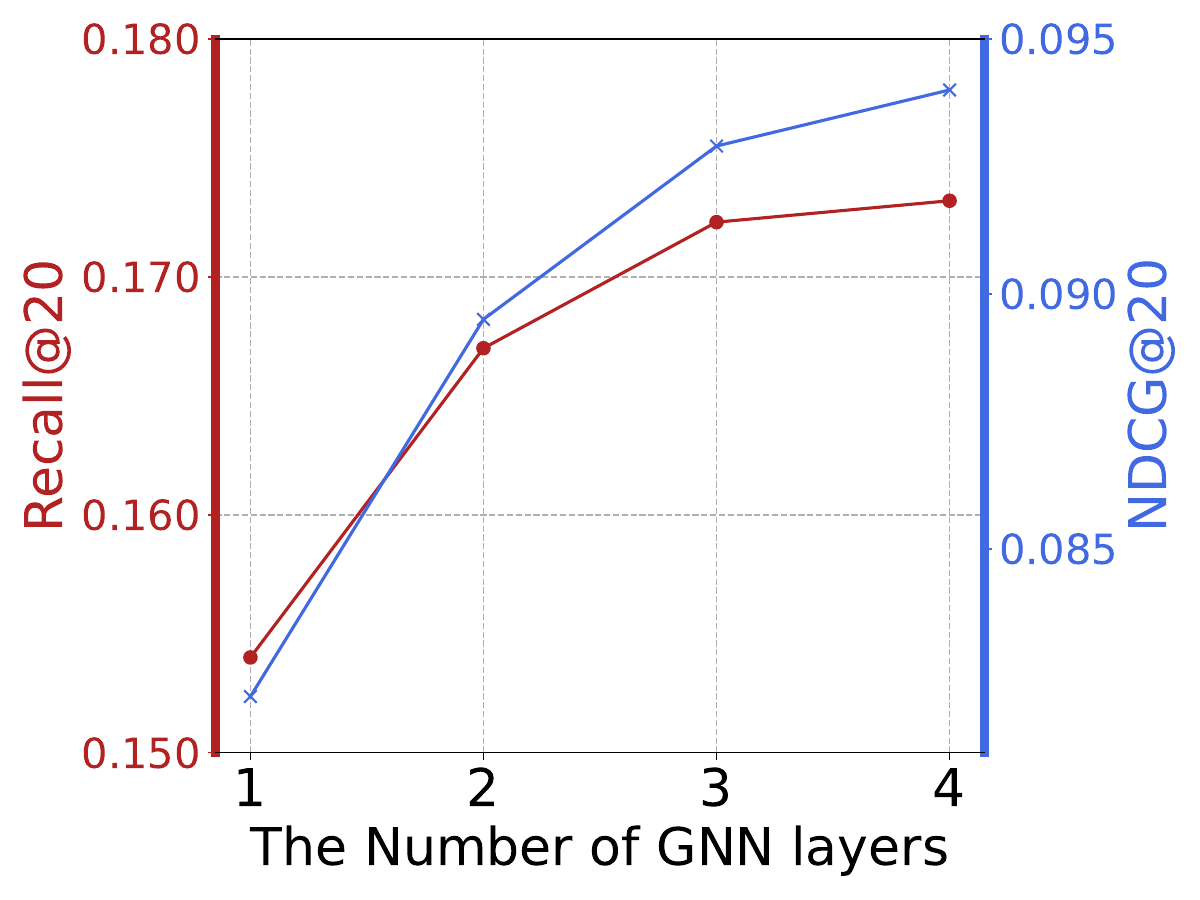}}
  \subfigure[MovieLens-20M]{\includegraphics[width=0.24\textwidth]{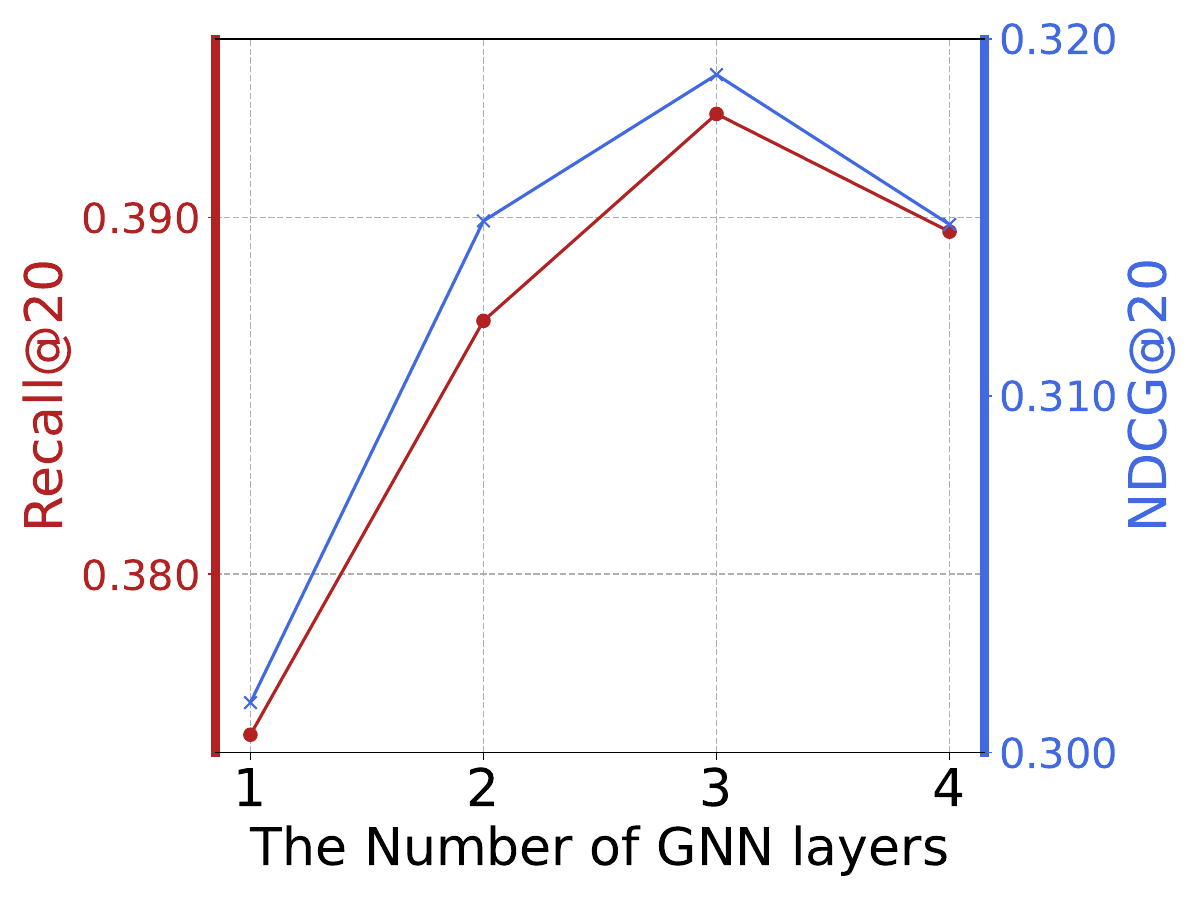}}
  \subfigure[Yelp2018]{\includegraphics[width=0.24\textwidth]{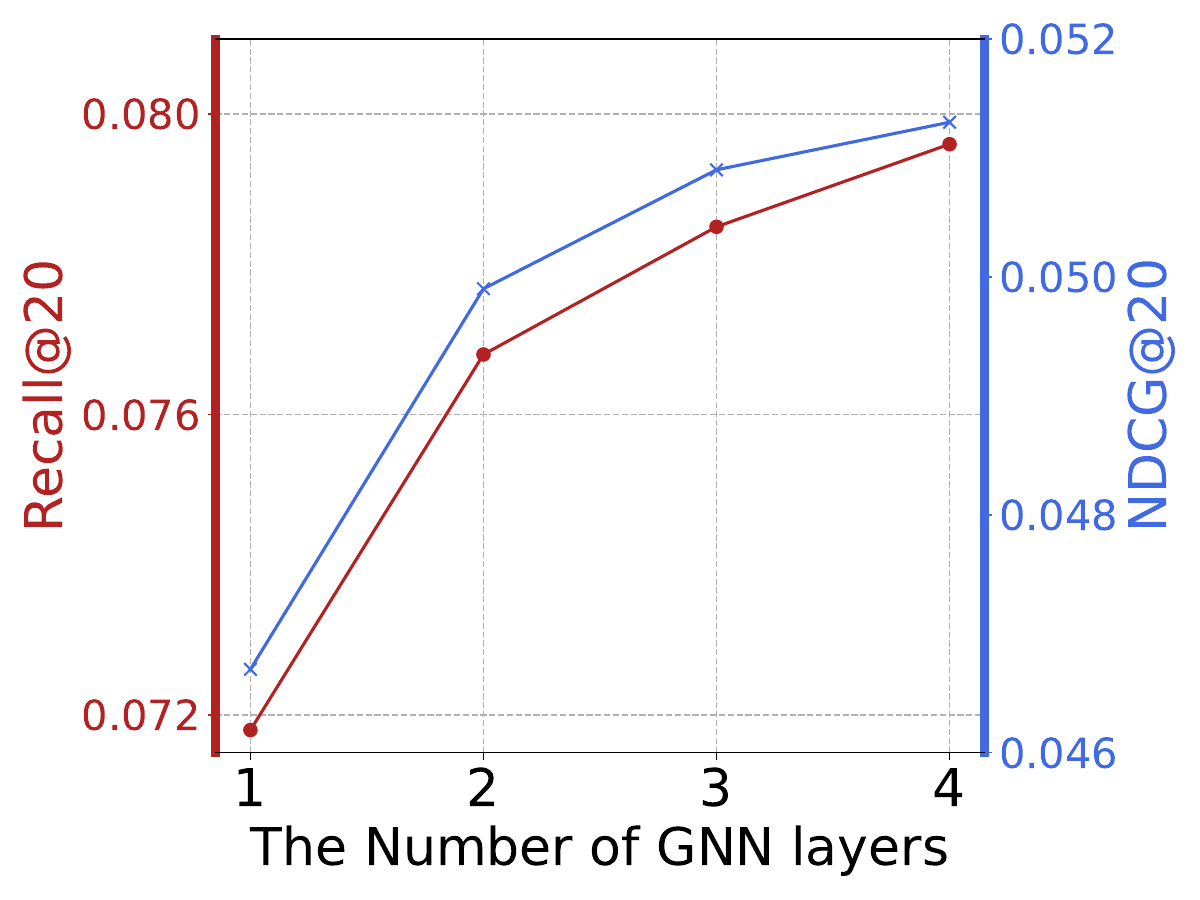}}
  \caption{The curves of Recall@$20$ and NDCG@$20$ when the number of GNN layers $L$ vary from 1 to 4 on four datasets.}
  \label{fig:layernum}
\end{figure}

 In the following experiments, we study the effects of the number of GNN layers $L$ and the temperature $\tau$ on the performance. For $L$, we vary its value in range of $\{1,2,3,4\}$ and demonstrate the corresponding performance in Fig. \ref{fig:layernum}. We have following findings: (1) The performance is inferior when $L=1$, indicating the necessity of involving high-order connectivity in both CG and KG into representations. (2) When depth is increased, the performance on some datasets consistently increases while goes up first and then decreases on the others. The phenomenon indicates high-order aggregation can bring more information as well as bring noise to representations.
 
For temperature $\tau$, we vary its value in range of $\{0.1, 0.4, 0.7, 1.0\}$ and conduct experiments on Last.FM and Yelp2018 dataset. From the results in Table \ref{tab:tau}, we learn that the in general the performance goes up when $\tau$ is increased on both datasets. The effect of $\tau$ on the performance of Yelp2018 dataset is relatively weaker than that of Last.FM dataset.

\begin{table}[h]
  \centering
  \caption{The performance with different temperature parameter $\tau$ on Last.FM and Yelp2018 datasets.}
  \label{tab:tau}
  \scalebox{0.8}{
  \begin{tabular}{l|ccc|ccc}
    \toprule[1pt]
    & \multicolumn{3}{c|}{Last.FM} & \multicolumn{3}{c}{Yelp2018}\\
    & AUC & Recall & NDCG & AUC & Recall & NDCG \\
    \hline
    \textbf{KDAR} & \textbf{0.9084} & \textbf{0.3991} & \textbf{0.2236} &  0.9594 & \textbf{0.0785} & \textbf{0.0509}\\
    \hline
    $\tau=0.1$ & 0.9003 & 0.3769 & 0.211 & 0.9593 & 0.0777 & 0.0504\\
    $\tau=0.4$ & 0.9048 & 0.3916 & 0.2204 & \textbf{0.9598} & 0.0777 & 0.0504\\
    $\tau=0.7$ & 0.9078 & 0.3948 & 0.2178 & \textbf{0.9598} & 0.0778 & 0.0504\\
    \bottomrule[1pt]
  \end{tabular}}
\end{table}

\subsection{Model Analysis(RQ4)}
In this section, we study the other benefits of our methods, including the performance on cold-start setting and long-tail item recommendations.
\begin{figure}
  \centering
  \subfigure[Last.FM]{\label{fig:cold_lfms}\includegraphics[width=0.24\textwidth]{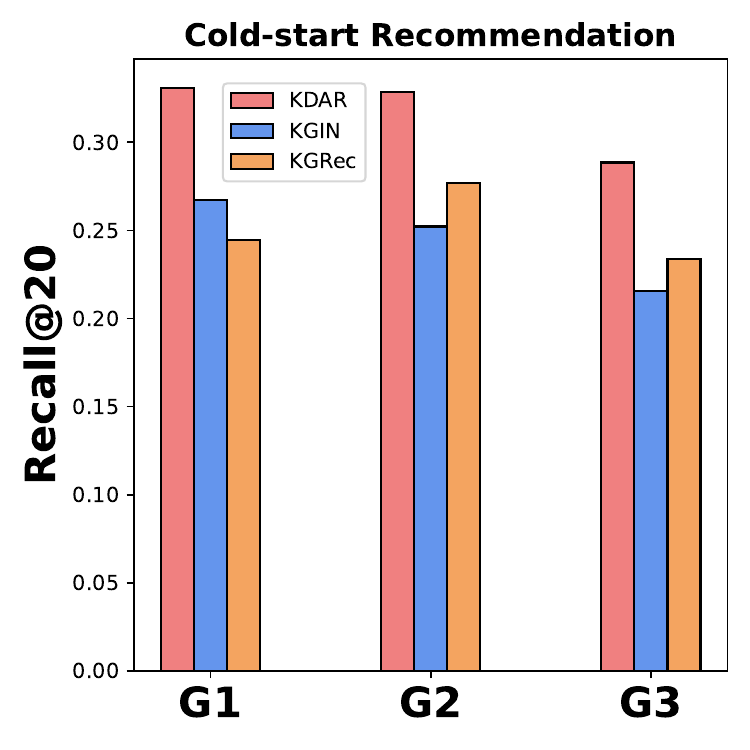}}
  \subfigure[Yelp2018]{\label{fig:cold_yelp}\includegraphics[width=0.24\textwidth]{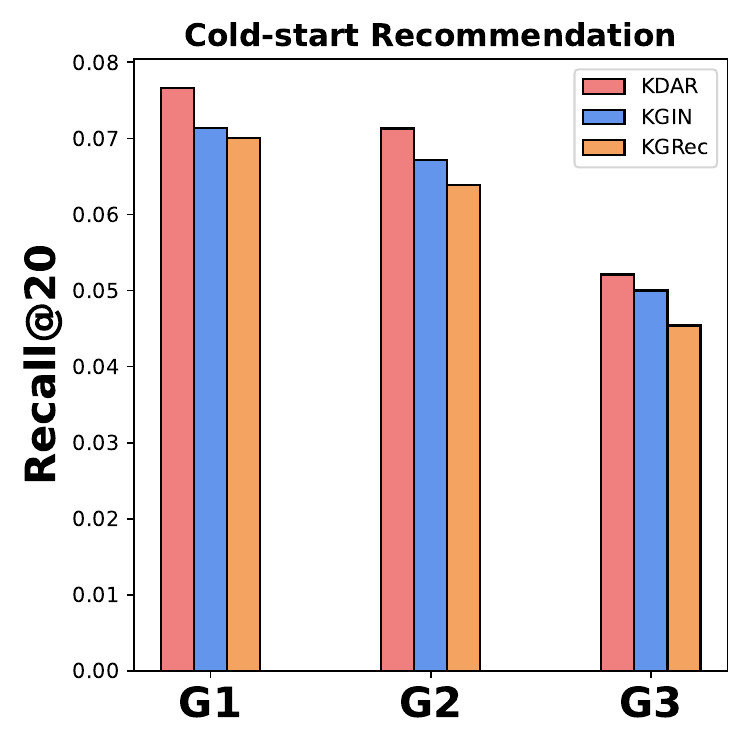}}
  \subfigure[Last.FM]{\label{fig:long_lfms}\includegraphics[width=0.24\textwidth]{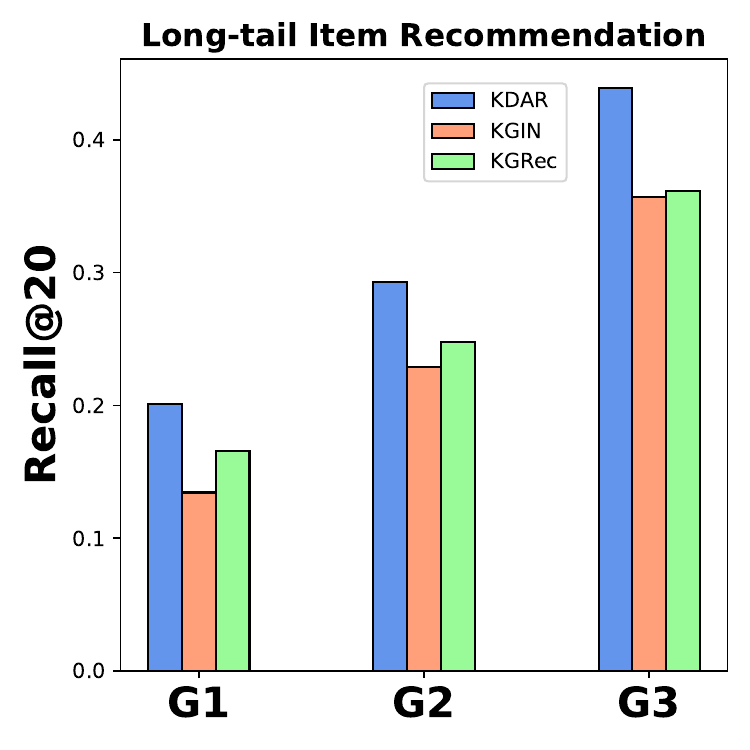}}
  \subfigure[Yelp2018]{
  \label{fig:long_yelp}
  \includegraphics[width=0.24\textwidth]{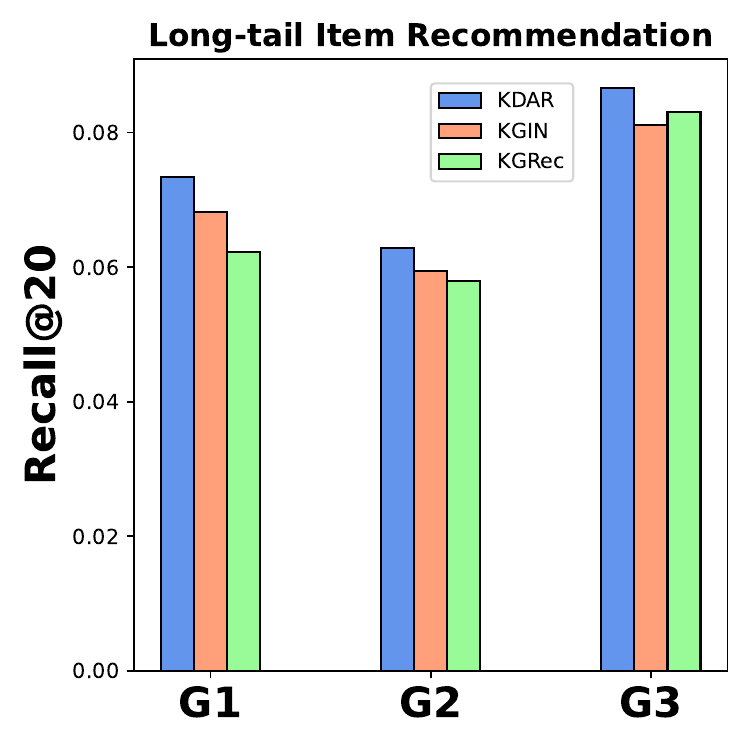}}
  \caption{The Recall@$20$ on cold-start setting (\ref{fig:cold_lfms} and \ref{fig:cold_yelp})and long-tail item recommendation (\ref{fig:long_lfms} and \ref{fig:long_yelp}) on Last.FM and Yelp2018 datasets.}
  
\end{figure}

\subsubsection{Cold-start Recommendation} We split the users in Last.FM and Yelp2018 datasets into three groups according to the number of their interactions, respectively. The smaller group number means the stronger cold-start effect. The results of our method and two baselines on each group are demonstrated the results in Fig. \ref{fig:cold_lfms} and \ref{fig:cold_yelp}. We can see that KDAR outperforms all baselines in each group, which demonstrates its superiority in alleviating cold-start problem at different levels. The results can be attributed to that the connection between user preferences and item attributes provides auxiliary to recommendation, thus alleviating the sparsity of collaborative data.

 \subsubsection{Long-tail Item Recommendation} To study the performance on long-tail items, we divide users into three groups according to the average occurrence numbers of their historical items. The results are illustrated in Fig. \ref{fig:long_lfms} and \ref{fig:long_yelp}. Our method outperforms all baselines. This can be attributed to that item representations are enhanced by attribute information, thus acquiring external knowledge to help recommendation making.

\section{Related Work}
In this section, we give a brief review of existed knowledge-aware recommendation methods. We divide them into four categories.

\subsection{KGE based Methods}
Methods of this category \cite{cke,ktup,kbe,mkr} take advantages of KGE techniques \cite{transe,transh,transr} to pre-train entity embeddings. CKE \cite{cke} learns item representations via TransR \cite{transr} and uses them to enhance matrix factorization (MF) \cite{mf}. KTUP \cite{ktup} utilizes TransH \cite{transh} to train a model for both KG completion and recommendation tasks. These methods learn entity embeddings based on transition constraints of graphs. However, this paradigm focuses on first-order connectivity of graphs, which ignores high-order connectivity.

\subsection{Path-based Methods}
This group of methods \cite{rkge,hete-mf,mcrec,herec} captures high-order connectivity of KG by selecting paths connecting user-item pairs via entities. The paths are fed into predictive models to provide extra knowledge. MCRec \cite{mcrec} explicitly incorporates path instances which are sampled via priority-based random walk based on meta-paths as the contexts in user-item interactions. Hete-MF \cite{hete-mf} generates similarity matrices from each meta-paths and incorporates them into matrix factorization. However, the sampling strategies  either may select low-quality (\textit{e.g.}, brute-force search) paths or require domain expert knowledge (\textit{e.g.}, designing meta-path patterns) and is poor in transferability.

\subsection{GNN-based Methods}
GNN-based methods \cite{kgcn,kgnnls,kgat,ckan,cgkgr,kgin} use aggregation scheme to capture information in KG. For each target entity, information of entities that are multi-hops away from it is integrated into it by graph aggregation. Thus both the first-order and high-order connectivity are modeled in GNN-based methods. KGAT \cite{kgat} combines KG and CG to form a holistic graph and employs an attention mechanism to discriminate the importance of the neighbors when performing aggregation on the graph. KGIN \cite{kgin} models the user-item interaction relation at user intent level and designs a relational-aware aggregation scheme, which improves the explainability of recommendation and captures the long-range semantics of relational paths.

\subsection{Contrastive Learning-based Methods}
Contrastive Learning-based methods \cite{kgcl,kgic,kgrec} introduce self-supervised learning scheme into GNN-based methods. Except the recommendation object, CL-based methods design special contrastive objects to improve the quality of representations. KGCL \cite{kgcl} adopts contrastive learning to reduce noise in KGs. KGRec \cite{kgrec} utilizes contrastive learning to align the KG representations with CF signals and learn the global rationale.

Our KDAR is built upon this category of work. The existed methods have flaws in fine-grained user preference modeling and further fall short in leveraging the preference-attribute connection to improve recommendation. Different from these works, our method proposes to enhance user and item side CF-based representations with attribute information and we design a multi-level collaborative alignment contrasting mechanism to better learn attribute-based representations.

\section{Conclusion, Future Work and Ethics Statement}
In this paper, we propose a novel graph CL-based KGR method named KDAR, which constructs attribute-based enhancing representations by attentive aggregation and utilizes contrastive learning to adjust the importance of attributes. We leverage the preference-attribute connection to make recommendations. Empirical studies are conducted to analyze effectiveness of KDAR. In the future, we will explore more complex methods to align contributions of attributes with the semantics of CF signals.

\textbf{Ethics Statement:} Our work aims to improve the accuracy of recommendations using processed data without leaking personal information. We believe our work is ethical.


%
%
%
\bibliographystyle{splncs04}
%
\bibliography{reference}





\end{document}